\documentclass[aps,pra,twocolumn,amssymb, amsfonts,amsmath,showpacs, superscriptaddress]{revtex4}
 \usepackage{graphicx}
  \usepackage{amssymb}
  \begin{document}
  
  \title{Simple Derivation of the Lindblad Equation}
\author{Philip Pearle}
\email{ppearle@hamilton.edu}
\affiliation{Department of Physics, Hamilton College, Clinton, NY  13323}
\date{\today}
\begin{abstract}
{The Lindblad equation is an evolution equation for the density matrix in quantum theory.  It is the general linear,  Markovian, form which ensures that the density matrix is hermitian, trace 1, positive and completely positive. Some elementary examples of the Lindblad equation are given. The derivation of the Lindblad  equation presented here is ``simple" in  that all it uses is the expression of a hermitian matrix in terms of its orthonormal eigenvectors and real eigenvalues.  Thus, it is appropriate for students who have learned the algebra of quantum theory. Where helpful, arguments are first given in a two-dimensional hilbert space. } 
\end{abstract}

\maketitle

\section{Introduction}

The density matrix ${\bf\rho}$ is a useful operator for quantum mechanical calculations.  For a given system, one may be unsure about what is the state vector.  If the possible state vectors and their associated probabilities are $\{|\psi_{i}\rangle, p_{i}\}$, one creates the \textit{proper}\cite{dEspagnat} density matrix 

\begin{equation}\label{Z1}
{\bf\rho}\equiv \sum_{i}p_{i}|\psi_{i}\rangle\langle\psi_{i}|.
\end{equation}

It is hermitian: ${\bf\rho}^{\dagger}={\bf\rho}$.  It is trace 1: $1= Tr{\bf\rho}\equiv \sum_{j}\langle \phi_{j}|{\bf\rho}|\phi_{j} \rangle$, where $\{|\phi_{j}\rangle\}$ are an arbitrary complete orthonormal set of vectors.  It is positive:  $\langle v|{\bf\rho}|v\rangle\geq 0$ for an arbitrary vector $ |v\rangle$.  All these properties can easily be verified from Eq.(\ref{Z1}).  

One can use the density matrix to conveniently calculate probabilities or mean values.  If a measurement is set up to result in one of the eigenstates  $|\phi_{i}\rangle$ of an operator ${\bf O}$, that outcome's probability is $\langle\phi_{i}|{\bf\rho}|\phi_{i}\rangle$ and the mean eigenvalue of ${\bf O}$ is $Tr{\bf O}{\bf\rho}$.

Because each individual state vector evolves unitarily under the system Hamiltonian {\bf H} (assumed for simplicity here to be time independent), $|\psi,t\rangle=\exp-i t{\bf H}|\psi,0\rangle$, the density matrix in Eq.(\ref{Z1}) satisfies the evolution equation 
\begin{equation}\label{Z2}
\frac{d}{dt}{\bf\rho}(t)=-i[{\bf H}, {\bf\rho}(t)]\implies{\bf\rho}(t)=e^{-i{\bf H}t}{\bf\rho}(0)e^{i{\bf H}t}.  
\end{equation}
\noindent  The operator acting on ${\bf\rho}(0)$ is often called a \textit{superoperator}\cite{Prig} since it describes a linear transformation on an operator:  it operates on both sides of $\rho$, so to speak.  

The case sometime arises where the system $S$ under consideration is a subsystem of a larger system $S+S'$, and  $S'$  is not measured. The pure (so-called because it is formed from a single state vector) density matrix for the joint system is
\[
\hbox{{\cal R}} (t)=\sum_{im}C_{im}(t)|\phi_{i}\rangle|\chi_{m}\rangle
\sum_{jn}C_{jn}(t)^{*}\langle\phi_{j}|\langle\chi_{n}|,
\]
\noindent where $\{|\phi_{i}\rangle\}$ $\{|\chi_{m}\rangle\}$ are orthonormal bases for $S$, $S'$ respectively, and $\sum_{im}|C_{im}(t)|^{2}=1$.
By taking the trace of $\hbox{{\cal R}}$ with respect to  $S'$, one arrives at an \textit{improper}\cite{dEspagnat} density matrix for $S$ from which predictions can be extracted:
\[
{\bf \rho} (t)=\sum_{ijm}C_{im}(t)C_{jm}(t)^{*}|\phi_{i}\rangle
\langle\phi_{j}|.
\]
One can easily see that this is hermitian, trace 1 and positive.  However, while the density matrix of $S+S'$ evolves unitarily, the density matrix of the subsystem $S$ evolving under the influence of  $S'$ generally does not evolve unitarily.  Nonetheless,  sometimes $d{\bf\rho}(t)/dt$ can be written in terms of ${\bf\rho}$ for a range of times earlier than $t$.  Sometimes that range is short compared to the time scale of evolution of  ${\bf\rho}$ so that one may make an approximation  whereby $d{\bf\rho}(t)/dt$ depends linearly just on ${\bf\rho}(t)$. This is quite useful, and is what shall be considered in this paper. 

In this case, the evolution equation is highly constrained by the requirements on ${\bf\rho}(t)$, to be satisfied at all times:  hermiticity, trace 1 and positivity. (The latter proves too general to simply implement, so a stronger requirement  is imposed, called complete positivity---see Section \ref{SecPos}).)  The result, for an $N$-dimensional Hilbert space, is the Lindblad\cite{Lind}  (or Lindblad-Gorini-Kossakowsky-Sudarshan\cite{Sud}) evolution equation for the density matrix:
\begin{eqnarray}\label{Z3}
&&\frac{d}{dt}{\bf\rho}(t)=-i[{\bf H},{\bf\rho}(t)]\nonumber\\
&&\quad-\frac{1}{2}\sum_{\alpha=1}^{N^{2}-1}[{\bf L}^{\alpha\dagger}{\bf L}^{\alpha}{\bf\rho}(t)
+{\bf\rho}(t){\bf L}^{\alpha\dagger}{\bf L}^{\alpha}-2{\bf L}^{\alpha}{\bf\rho}(t){\bf L}^{\alpha\dagger}].
\nonumber\\
\end{eqnarray}
In Eq.(\ref{Z3}), the hamiltonian ${\bf H}$ is an arbitrary hermitian operator, but the Lindblad operators $\{{\bf L}^{\alpha}\}$ are completely arbitrary operators.  

Actually, as shall be shown, there need be no limitation on the number of terms in the sum in Eq.(\ref{Z3}),  but this can always be reduced to a sum of 
$N^{2}-1$ terms.  

 It is not a necessary condition, but if the equation is to be time-translation-invariant, the operators are time-independent. 

\section{Lindblad Examples}

Before deriving Eq.(\ref{Z3}), we give a few examples of the non-unitary evolutions  it describes.   Since unitary evolution is well known, we shall let ${\bf H}=0$.  It shall be seen that relaxation to some equilibrium (constant) density matrix is readily described. 

For simplicity, four of the five examples shall be in a $N=2$ hilbert space (a restriction that is readily lifted).  

\subsection{Random Phases}
Consider a state vector written in a basis whose phase factors undergo random walk. 

Given an initial state vector $|\psi\rangle=a|\phi_{1}\rangle+b|\phi_{2}\rangle$ ($|a|^{2}+|b|^{2}=1$), suppose at time $t$, it has evolved to  
\[
|\psi,t\rangle=ae^{i\theta_{1}}|\phi_{1}\rangle+be^{i\theta_{2}}|\phi_{2}\rangle
\]
\noindent    with probability 
\[
P(\theta_{1}, \theta_{2})d\theta_{1}d \theta_{2}=\frac{d\theta_{1}}{\sqrt{2\pi\lambda_{1}t}}\frac{d\theta_{1}}{\sqrt{2\pi\lambda_{2}t}}e^{-\frac{1}{2\lambda_{1}t}\theta_{1}^{2}}e^{-\frac{1}{2\lambda_{2}t}\theta_{2}^{2}}. 
\] 
The density matrix is
\begin{eqnarray}
{\bf \rho}(t)&=&\int_{-\infty}^{\infty} \int_{-\infty}^{\infty}d\theta_{1}d\theta_{2}P(\theta_{1}, \theta_{2})
|\psi\rangle\langle\psi|\nonumber\\
&=&|a|^{2}|\phi_{1}\rangle\langle\phi_{1}|+|b|^{2}|\phi_{2}\rangle\langle\phi_{2}|\nonumber\\
&&+e^{-\frac{1}{2}t[\lambda_{1}+\lambda_{2}]}\big[ab^{*}|\phi_{1}\rangle\langle\phi_{2}|+a^{*}b|\phi_{2}\rangle\langle\phi_{1}|\big]. \nonumber
\end{eqnarray}
\noindent  We see that the off-diagonal elements decay at a fixed rate while the diagonal elements remain constant.  It satisfies 
\[
\frac{d}{dt}{\bf \rho}(t)=-\frac{1}{2}[\lambda_{1}+\lambda_{2}]\Big[{\bf \rho}(t)-\sum_{i=1}^{2}{\bf Q}_{i}{\bf \rho}(t){\bf Q}_{i}\Big],
\]
\noindent where the projection operator ${\bf Q}_{i}\equiv|\phi_{i}\rangle\langle\phi_{i}|$.  To see that this is a Lindblad equation, note that  
${\bf Q}_{i}^{\dagger}={\bf Q}_{i}$ and $\sum_{i=1}^{2} {\bf Q}_{i}^{\dagger}{\bf Q}_{i}={\bf 1}$. There are two Lindblad operators:  identify ${\bf L}^{i}\equiv\sqrt{(\lambda_{1}+\lambda_{2})}{\bf Q}_{i}$ 
in Eq.(\ref{Z3}).

\subsection{Unitary Jump}

Suppose in time $dt$, a state vector $|\psi, t\rangle$ has probability $\lambda dt$ of changing to $\exp-i{\bf G}|\psi, t\rangle$ (probability $1-\lambda dt$ of being unchanged), where 
${\bf G}$ is a hermitian operator and $\exp-i{\bf G}\neq{\bf 1}$.  The density matrix at time $t+dt$ is therefore           
\[
	{\bf \rho}(t+dt)=(1-\lambda dt){\bf \rho}(t)+\lambda dte^{-i{\bf G}}{\bf \rho}(t)e^{i{\bf G}}, 
\]
\noindent so its evolution equation is  
\[
		\frac{d}{dt}{\bf \rho}(t)=-\lambda[{\bf \rho}(t)-e^{-i{\bf G}}{\bf \rho}(t)e^{i{\bf G}}].
\]
\noindent This is of the Lindblad form, with one Lindblad operator ${\bf L}\equiv  \sqrt{2\lambda}\exp{-i{\bf G}}$. 

In the basis where ${\bf G}$ is diagonal with elements $(g_{1},g_{2})$, we get 
$d\rho_{ii}/dt=0$ and
\[
\frac{d}{dt}\rho_{12}(t)=-\lambda\rho_{12}(t)\Big[1-e^{i(g_{2}-g_{1})}\Big]. 
\]
\noindent So, again, its diagonal elements remain constant.   Its off-diagonal elements  decay at the fixed rate $\lambda[1-\cos(g_{2}-g_{1})]$ 
and their phases change. 
 
\subsection{Random Unitary Transformation}
Suppose in time $dt$ a state vector $|\psi,t\rangle$ undergoes a unitary transformation to
\[ 
	e^{-i{\bf G}\theta}|\psi,t\rangle=[1-i{\bf G}\theta -\frac{1}{2}({\bf G}\theta)^{2}+ ...|\psi,t\rangle
\]
 with probability 
\[
P(\theta)d\theta = \frac{d\theta}{\sqrt{4\pi\lambda dt}}e^{-\frac{1}{4\lambda dt}\theta^{2}}. 
\] 
\noindent  The density matrix at $t+dt$ (neglecting terms of order higher than $dt$) is given by
\begin{eqnarray}
&&{\bf \rho}(t+dt)=\int_{-\infty}^{\infty}P(\theta)d\theta e^{-i{\bf G}\theta}{\bf \rho}(t)e^{-i{\bf G}\theta}\nonumber\\
&&=\int_{-\infty}^{\infty}P(\theta)d\theta\Bigg[{\bf \rho}(t)-\theta^{2}\Big[\frac{1}{2}{\bf G}^{2}{\bf \rho}(t)+{\bf \rho}(t)\frac{1}{2}{\bf G}^{2}-{\bf G}{\bf \rho}(t){\bf G}\Big]\Bigg]\nonumber\\
&&={\bf \rho}(t)-\frac{\lambda dt}{2}\Big[{\bf G}^{2}{\bf \rho}(t)+{\bf \rho}(t){\bf G}^{2}-2{\bf G}{\bf \rho}(t){\bf G}\Big]\nonumber
\end{eqnarray}
\noindent giving the Lindblad equation
\[
\frac{d}{dt}{\bf \rho}(t)=-\frac{\lambda}{2}\Big[{\bf G}^{2}{\bf \rho}(t)+{\bf \rho}(t){\bf G}^{2}-2{\bf G}{\bf \rho}(t){\bf G}\Big]=-\frac{\lambda}{2}[{\bf G}[{\bf G}, {\bf \rho}(t) ]
\]
\noindent with one Lindblad operator ${\bf L}\equiv \sqrt{\lambda} {\bf G}$.

In the basis where ${\bf G}$ is diagonal with elements $(g_{1},g_{2})$, 
\[
\frac{d}{dt}\rho_{ij}(t)=-\frac{\lambda}{2}(g_{i}-g_{j})^{2}\rho_{ij}(t). 
\]
\noindent So, again, its diagonal elements remain constant but its off-diagonal elements decay at a rate determined by the difference in eigenvalues. 

\subsection{State Exchange}
Suppose, in time $dt$, with probability $\lambda dt$, a state vector $|\psi, t\rangle$  exchanges its basis states $|\phi_{1}\rangle, |\phi_{2}\rangle$, becoming ${\bf\sigma}|\psi, t\rangle$, where  
$\langle\phi_{i}| {\bf \sigma}|\phi_{j}\rangle$ is the Pauli matrix with diagonal elements 0 and off-diagonal elements 1.    It is easy to see that the density matrix evolution equation is 

\[
\frac{d}{dt}{\bf \rho}(t)=-\lambda\Big[{\bf \rho}(t)-{\bf\sigma} {\bf \rho}(t){\bf\sigma}\Big]
\] 
\noindent and  is of the Lindblad form, with one Lindblad operator ${\bf L}\equiv \sqrt{2\lambda}{\bf\sigma}$.

The density matrix elements therefore satisfy
\begin{eqnarray}
\frac{d}{dt} \rho_{11}(t)&=&-\frac{d}{dt} \rho_{22}(t)=- \lambda[\rho_{11}(t)- \rho_{22}(t)],\nonumber\\
\frac{d}{dt} \rho_{12}(t)&=&-\frac{d}{dt} \rho_{21}(t)=- \lambda[\rho_{12}(t)- \rho_{21}(t)].\nonumber
\end{eqnarray}
\noindent The diagonal density matrix elements change in this example, decaying to 1/2.  The off-diagonal matrix elements keep their real parts while the imaginary parts decay to 0.

\subsection{State Transitions}
Here we consider arbitrary $N$. Suppose in time $dt$, a state vector $|\psi, t\rangle$  makes a transition to state  
\[
|m\rangle \frac{\langle n| \psi, t\rangle}{|\langle n| \psi, t\rangle|}
\]
\noindent with probability
\[
P_{mn}\equiv p_{m} \lambda dt|\langle \psi, t| n\rangle|^{2}
\]
\noindent ($p_{m}\geq 0$,  $ \sum_{m=1}^{N}p_{m}=1$).  The probability of all such transitions is $\sum_{mn}P_{mn}=\lambda dt$, so the state vector is unchanged with probability $1-\lambda dt$.  Define  
${\bf Q}_{mn}\equiv | m\rangle \langle n|$.   Note that 
$\sum_{m,n=1}^{N}p_{m}{\bf Q}_{mn}^{\dagger}{\bf Q}_{mn}={\bf 1}$.

The density matrix at time $t+dt$ is
\[
{\bf \rho}(t+dt)=(1-\lambda dt){\bf \rho}(t)+\lambda dt\sum_{m,n=1}^{N}p_{m}{\bf Q}_{mn} {\bf \rho}(t){\bf Q}_{nm}, 
\]
\noindent so its evolution equation is 
\[
\frac{d}{dt}{\bf \rho}(t)=-\lambda\Big[{\bf \rho}(t)-\sum_{m,n=1}^{N}p_{m}{\bf Q}_{mn} {\bf \rho}(t){\bf Q}_{nm}\Big].
\]
\noindent This is of the Lindblad form, with ${\bf N}^{2}$ (one more than the necessary maximum!) Lindblad operators 
${\bf L}^{mn}\equiv\sqrt{2\lambda p_{m}}{\bf Q}_{mn}$: 

The matrix elements of the density matrix obey
\begin{eqnarray}
\frac{d}{dt} \rho_{rs}(t)&=&-\lambda\Big[ \rho_{rs}(t)-p_{r}\delta_{rs}\sum_{n}\rho_{nn}(t)\Big]\nonumber\\
&=&-\lambda\Big[ \rho_{rs}(t)-p_{r}\delta_{rs}\Big].\nonumber
\end{eqnarray}
\noindent The off-diagonal elements decay at a uniform rate.  The diagonal elements do not remain constant. They decay to predetermined values $p_{r}$:
\[
\rho_{rr}(t)=\rho_{rr}(0)e^{-\lambda t}+p_{r}\Big[1-e^{-\lambda t}\Big].
\]
This might be useful in modeling the  approach to thermal equilibrium, where the states $|m\rangle$ are energy eigenstates and $p_{r}$ is the Boltzmann probability $Z^{-1}\exp-E_{r}/kT$.

  \section{Application of Constraints} 
 
 We now turn to deriving the Lindblad equation as the most general equation satisfying the constraints.  
 
 While the hilbert space discussed here shall 
 be assumed of dimension $N$,  $N$ may be allowed to go to to infinity and, also, the argument may readily be extended to a continuum basis. 
 
To make the argument  
easier to follow,  examples of how its steps apply to a two-dimensional hilbert space shall occasionally be inserted.  
 
 The  \textit{Markov} constraint is that the density matrix ${\bf\rho}(t')\equiv{\bf\rho}'$ at a later time $t'$,  depends only upon the density matrix ${\bf\rho}(t)\equiv{\bf\rho}$ at an earlier time $t$,  not upon the density matrix over a range of earlier times. 
 
The \textit{linearity} constraint, combined with the Markov constraint, is that the matrix elements of ${\bf\rho}'$ can be written as the sum of constants multiplying the matrix elements of ${\bf\rho}$  rather than, say, powers of the matrix elements of ${\bf\rho}$ or any other kind of function of these matrix elements: 
 \begin{equation}\label{X1}
\rho'_{ij}=\sum_{r,s=1}^{N}A_{ir,js}\rho_{rs}.  
\end{equation}
\noindent  Here $\rho'_{ij}\equiv\langle\phi_{i}|{\bf\rho}' |\phi_{j}\rangle$  with $ |\phi_{j}\rangle$  some convenient orthonormal basis and,  similarly, ${\bf\rho}$ is expressed in the same basis.  The constants $A_{ir,js}$ can be functions of $t'$, $t$. There are $N^{4}$ constants, and each can be complex, so there are $2N^{4}$ real constants involved 
in Eq.(\ref{X1}). 

The \textit{hermiticity} constraint $\rho_{ij}^{'\dagger}\equiv\rho_{ji}^{'*}=\rho'_{ij}$, applied to  Eq.(\ref{X1}),  results in 
\begin{equation}\label{X2}
\sum_{r,s=1}^{N}[A_{js,ir}^{*}-A_{ir,js}]\rho_{rs}.
\end{equation}

\subsubsection{Two-dimensional space: hermiticity}

Suppose we have the equation
\[
B_{11}\rho_{11}+B_{12}\rho_{12}+B_{21}\rho_{21}+B_{22}\rho_{22}=Tr{\bf B}{\bf \rho}=0,
\] 
\noindent (the $B_{ij}$ are constants) which holds for all possible density matrices.  Then, one can see ${\bf B}=0$  as follows. 

First choose 
the density matrix $\rho_{11}=1$, with all other elements vanishing: thus,  $B_{11}=0$.  Similarly, one shows $B_{22}=0$. Next, employ the density matrix
$ \rho_{ij}=1/2$, which results in $B_{12}+B_{21}=0$  Finally, use the density matrix $\rho_{11}=\rho_{22}=1/2$, $\rho_{12}=-\rho_{21}=i/2$, which results in 
$B_{12}-B_{21}=0$ and so $B_{12}=B_{21}=0$. Therefore, ${\bf B}=0$

The four density matrices used here, 
\[
\frac{1}{2}({\bf 1}+ {\bf \sigma}^{3}), \medspace \frac{1}{2}({\bf 1}- {\bf \sigma}^{3}), \medspace \frac{1}{2}({\bf 1}+{\bf \sigma}^{1}),\medspace \frac{1}{2}({\bf 1}+{\bf \sigma}^{2}),
\]
\noindent (written in terms of the Pauli matrices)
we shall call \textit{the density matrix basis}.   Any $2\times2$ matrix can be written as a linear sum with constant (complex) coefficients of these four matrices. 
                               More than that, they form a \textit{matrix basis for hermitian matrices}, in that any hermitian matrix can be written as a linear sum with constant (real) coefficients of these four matrices. More than that, and this is the reason for their deployment here, they form a \textit{matrix basis for density matrices}, in that any density matrix can be written as a linear sum with constant positive real coefficients of these four matrices such that the sum of the coefficients add up to 1.  
                               
                               This basis is to be distinguished from another basis,  the Pauli matrices plus the identity matrix, which we shall call the \textit{Pauli+1} basis. This is also a matrix basis for hermitian matrices but it is \textit{not}   a density matrix basis.

\subsubsection{N-dimensional space: hermiticity}

 Generalizing, if we have an equation  
\begin{equation}\label{X3}
\sum_{r,s=1}^{N}B_{sr}\rho_{rs}\equiv Tr{\bf B\rho}=0
\end{equation}
\noindent  for a matrix ${\bf B}$, which holds for all valid ${\bf\rho}$, then ${\bf B}=0$.

This can be seen by using an $N^{2}$-size density matrix  basis,  (generalizing the $2^{2}$-size density matrix basis of the previous section).  First choose $\rho_{kk}=1$ with all other elements vanishing,   which implies $B_{kk}=0$. Then  for particular values of $k,l$, choose $\rho_{kk}=\rho_{ll}=\rho_{kl}=\rho_{lk}=1/2$ with all other elements vanishing, from which one finds $B_{kl}+B_{lk}=0$. Finally,  choose $\rho_{kk}=\rho_{ll}=i\rho_{kl}=-i\rho_{lk}=1/2$,  from which one  finds $B_{kl}-B_{lk}=0$,   so $B_{kl}=B_{lk}=0$.  Letting $k$, $l$ range over all possible pairs of indices results in ${\bf B}=0$. 

\subsubsection{Two-dimensional space: evolution equation and trace constraint}

It therefore follows from Eq.(\ref{X2}) that 
\[
A_{js,ir}^{*}=A_{ir,js}
\]
\noindent where each index can take on the values 1 or 2. 

A matrix ${\bf B}$ for which $B_{m, n}^{*}=B_{n, m}$ is a hermitian matrix.  Therefore, ${\bf A}$ is a hermitian matrix, where we regard the number pairs  11, 12, 21, 22  as 
four different indices.  That is,  ${\bf A}$ is a $4\times4$ dimensional matrix.  The most general $4\times4$ hermitian  matrix is characterized by 16 real numbers (the four real diagonal matrix elements and the six complex matrix elements above the diagonal).  Since there are 32 real numbers which characterized the most general superoperator in a two dimensional space, the condition of hermiticity of the density matrix has cut that number in half.  

A hermitian matrix can be written in terms of its orthonormal eigenvectors and eigenvalues, and that decomposition shall prove very useful here.  There are four real eigenvalues, 
$\lambda^{\alpha}$, where $\alpha =1, 2, 3, 4$.  Corresponding to each eigenvalue is an eigenvector ${\bf E}^{\alpha}$ in the four dimensional complex vector space.  

The four complex components of
$E_{ir}^{\alpha}$ make 32 real numbers, but they are constrained. Each eigenvector is normalized to 1: $\sum_{i,r=1}^{2}E^{*\alpha}_{ir}E^{\alpha}_{ir}=1+i0$ provides 8 constraints, lowering the number of free components to 24. The orthogonality of ${\bf E}^{1}$ to the other three vectors provides 6 constraints, the orthogonality of ${\bf E}^{2}$ to the remaining two vectors provides 4 constraints, and the orthogonality of ${\bf E}^{3}$ to ${\bf E}^{4}$ provides two constraints. Thus, there are 12 constraints on the 24 free components, so the eigenvectors contain 12 free components.  These, together with the four eigenvalues, comprise the 16 real numbers characterizing ${\bf A}$.

An example of such an orthonormal basis is given by $1/\sqrt{2}$ multiplying the  Pauli+1 basis.    If we write the four components 
of ${\bf E}^{\alpha}$ as a four dimensional vector with components $[E_{11}^{\alpha}, E_{12}^{\alpha},E_{21}^{\alpha},E_{22}^{\alpha}]$, then 
 ${\bf E}^{1}\equiv2^{-1/2}{\bf \sigma}^{1}$ has components $2^{-1/2}[0,1,1,0]$, ${\bf E}^{2}\equiv2^{-1/2}{\bf \sigma}^{2}$ has components $2^{-1/2}[0,-i,i,0]$, ${\bf E}^{3}\equiv2^{-1/2}{\bf \sigma}^{3}$ has components $2^{-1/2}[1,0,0,-1]$, ${\bf E}^{4}\equiv2^{-1/2}{\bf 1}$ has components $2^{-1/2}[1,0,0,1]$,  It is easy to verify that this is an orthonormal set of vectors. 

Although each ${\bf E}^{\alpha}$ is a vector in a four dimensional space, with  four components  $E_{ij}^{\alpha}$,  
 ${\bf E}^{\alpha}$ can also 
be regarded as an operator in the two-dimensional hilbert  space with four matrix elements $E_{ij}^{\alpha}$.  This leads to a neat way of writing the orthogonality relations for these eigenvectors. Instead of 
$\sum_{i=1}^{N}\sum_{r=1}^{N}E_{ir}^{\alpha}E_{ir}^{\beta*}=\delta^{\alpha\beta}$, we can write 
\[
Tr{\bf E}^{\alpha}{\bf E}^{\beta\dagger} =\delta^{\alpha\beta},
\]
\noindent where ${\bf E}^{\beta\dagger}$ is the hermitian conjugate (complex conjugate transpose) of ${\bf E}^{\beta}$.  It is easy to see how 
this works for the example where ${\bf E}^{\alpha}$ is $1/\sqrt{2}\times$ the Pauli+1 basis.

The expression for the components of ${\bf A}$ written in terms of its eigenvectors and eigenvalues is 
\[
A_{ir,js}=\sum_{\alpha=1}^{4}\lambda^{\alpha}E^{\alpha}_{ir}E^{*\alpha}_{js}.
\]
\noindent Putting this into Eq.(\ref{X1}) results in the \textit{evolution equation}
\[
\rho'_{ij}=\sum_{\alpha=1}^{4}\lambda^{\alpha}\sum_{r,s=1}^{2}E^{\alpha}_{ir}\rho_{rs}E^{*\alpha}_{js}=\sum_{\alpha=1}^{4}\lambda^{\alpha}{\bf E}^{\alpha}{\bf\rho} {\bf E}^{\alpha\dagger}.  
\]

	Now, lets impose the \textit{trace  constraint}, i.e., $\sum_{i=1}^{2}\rho'_{ii}=1$.  In terms of components this says
\[
1=\sum_{\alpha=1}^{4}\lambda^{\alpha}\sum_{i=1}^{2}\sum_{r,s=1}^{2}E^{\alpha}_{ir}\rho_{rs}E^{*\alpha}_{is}.
\] 
\noindent Writing $1=\sum_{i=1}^{2}\rho_{ii}=\sum_{r,s=1}^{2}\delta_{sr}\rho_{rs}$, the trace  constraint can be written as 
\[
\sum_{r,s=1}^{2}\Bigg[\sum_{\alpha=1}^{4}\lambda^{\alpha}\sum_{i=1}^{2}E^{\alpha\dagger}_{si}E^{\alpha}_{ir}-\delta_{sr}\Bigg]\rho_{rs}=0
\]
\noindent or in matrix notation as
\[
Tr\Bigg[\sum_{\alpha=1}^{4}\lambda^{\alpha}{\bf E}^{\alpha\dagger}\bf E^{\alpha}-{\bf 1}\Bigg]{\bf \rho}=0.
\]
\noindent where ${\bf 1}$ is the unit matrix. This must hold for arbitrary ${\bf \rho}$.  We have seen how to handle 
such an expression.  By successively putting in the four density basis matrices, we obtain the trace constraint
\[
\sum_{\alpha=1}^{4}\lambda^{\alpha}{\bf E}^{\alpha\dagger}\bf E^{\alpha}=\bf 1.
\] 

\subsubsection{N-dimensional space: trace constraint}

The $N$-dimensional case works just like the two-dimensional case. It  follows from Eq.(\ref{X2}) that ${\bf A}$ can be viewed as  an $N^{2}\times N^{2}$ hermitian matrix.  It has $N^{2}$ real eigenvalues. Its  $N^{2}$  complex eigenvectors $E_{ir}^{\alpha}$ satisfy the orthonormality conditions

\begin{equation}\label{X4a}
\sum_{i=1}^{N}\sum_{r=1}^{N}E_{ir}^{\alpha}E_{ir}^{\beta*}
=Tr{\bf E}^{\alpha}{\bf E}^{\beta\dagger} =\delta^{\alpha\beta}.
\end{equation}
\noindent With {\bf A}  written in terms of its eigenvalues and eigenvectors, Eq.(\ref{X1}) becomes the evolution equation 
\begin{eqnarray}\label{X5}
\rho'_{ij}&=&\sum_{\alpha=1}^{N^{2}}\lambda^{\alpha}\sum_{r,s=1}^{N}E_{ir}^{\alpha}E_{js}^{\alpha*}\rho_{rs}\quad\hbox{or}\nonumber\\
 {\bf \rho}'&=&\sum_{\alpha=1}^{N^{2}}\lambda^{\alpha}{\bf E}^{\alpha}{\bf\rho} {\bf E}^{\alpha\dagger}.
\end{eqnarray}
\noindent  $\lambda^{\alpha}$ and ${\bf E}^{\alpha}$ depend upon $t'-t$, but we shall not write that dependence until it is needed.

	Next, imposition of the trace constraint on Eq.(\ref{X5}),  with  $Tr{\bf \rho}'=1=Tr{\bf 1\rho}$,  gives
\[
Tr\Big[\sum_{\alpha=1}^{N^{2}}\lambda^{\alpha}{\bf E}^{\alpha\dagger}{\bf E}^{\alpha}-{\bf 1}\Big]{\bf \rho} =0.
\]
\noindent Using the density matrix basis as in Eq.(\ref{X3}) et seq., we obtain the trace constraint: 
\begin{equation}\label{X6}
\sum_{\alpha=1}^{N^{2}}\lambda^{\alpha}{\bf E}^{\alpha\dagger}{\bf E}^{\alpha}={\bf 1}.
\end{equation}

By taking the trace of Eq.(\ref{X6}) and using Eq.(\ref{X4a}), we find the interesting relation
\[
\sum_{\alpha=1}^{N^{2}}\lambda^{\alpha}=N.
\]
\section{Complete Positivity}\label{SecPos}

The final constraint is \textit{positivity}.  This says, given an arbitrary N-dimensional vector $|v\rangle$, that the expectation value of the density matrix ${\bf \rho}'$ is non-negative.  
This constraint, applied to Eq.(\ref{X5}), is
\begin{equation}\label{X7}
0\leq\langle v|{\bf \rho}'|v\rangle=\sum_{\alpha=1}^{N^{2}}\lambda^{\alpha}\langle v|{\bf E}^{\alpha}{\bf\rho} {\bf E}^{\alpha\dagger}|v\rangle=\sum_{\alpha=1}^{N^{2}}\lambda^{\alpha}
\langle v_{\alpha}|{\bf \rho}|v_{\alpha}\rangle,
\end{equation}
where we have defined ${\bf E}^{\alpha\dagger}|v\rangle\equiv|v_{\alpha}\rangle$. 

Positivity of ${\bf\rho}$ ensures  $\langle v_{\alpha}|{\bf \rho}|v_{\alpha}\rangle\geq 0$.  Thus, we see from Eq.(\ref{X7}), if all the  $\lambda^{\alpha}$'s are non-negative, then ${\bf \rho}'$ will be positive too. 

However  $\lambda^{\alpha}\geq 0$, while just shown to be \textit{sufficient} for ${\bf \rho}'$ to be positive, is not \textit{necessary}.  In the next section, we shall give an example where an eigenvalue is negative, yet  ${\bf\rho}'$ is positive!

Therefore, a stronger condition than positivity is necessary to ensure that  $\lambda^{\alpha}\geq 0$.  This condition, presented after the example, is \textit{complete positivity}. 

\subsubsection{Two-dimensional space: example of a positive density matrix with negative eigenvalue}

This example uses the Pauli+1 eigenvectors ${\bf E}^{\alpha}=2^{-1/2}{\bf \sigma}^{\alpha}$ and $2^{-1/2}{\bf 1}$.  (Note that the trace constraint (\ref{X6})  is  satisfied, provided $\sum_{\alpha=1}^{4}\lambda^{\alpha}=2$, since the square of each of the Pauli+1 
matrices is  $2^{-1/2}{\bf 1}$.)  Choose  $\lambda^{1}=\lambda^{2}
=-\lambda^{3}=\lambda^{4}=1$:  

\[
{\bf \rho}'
=\frac{1}{2}[{\bf \sigma}^{1}{\bf\rho} {\bf \sigma}^{1}+{\bf \sigma}^{2}{\bf\rho} {\bf \sigma}^{2}-{\bf \sigma}^{3}{\bf\rho} {\bf \sigma}^{3}+{\bf1}{\bf\rho} {\bf 1}]
=\begin{bmatrix}\rho_{22}&\rho_{12}\\\rho_{21}&\rho_{11}
\end{bmatrix}.
\]
\noindent  ${\bf \rho}'$ is just ${\bf \rho}$ with its diagonal elements exchanged.  Thus,  because ${\bf \rho}$ is positive, then ${\bf \rho}'$ is  positive. This is
a particularly simple example of a more general case discussed in Appendix {\ref A}. 

\subsubsection{N-dimensional space: definition of complete positivity}

It is  not positivity but, rather, \textit{complete positivity} that makes the non-negative eigenvalue condition necessary. Here is what it means.

Add to our system a non-interacting and non-evolving additional system in its own $N$-dimensional hilbert space. The enlarged hilbert space is of dimension $N^{2}$. The simplest state vector in the enlarged space is a direct product $|\phi_{i}\rangle |\chi_{j}\rangle$: $|\phi_{i}\rangle$ is a vector from the original hilbert space, $|\chi_{j}\rangle$ is a vector from the added system.  The general state vector in the joint space is the sum of such products with c-number coefficients.  

Form an arbitrary density matrix ${ \cal R}$ for the enlarged system. Suppose it evolves according to Eq.(\ref{X5}), where ${\bf E}^{\alpha}$ is replaced by  ${\bf E}^{\alpha}\times1$ (i.e., the evolution has no effect on the vectors of the added system.) Complete positivity says that the resulting density matrix ${\cal R}'$ must be positive. 

\subsubsection{Two-dimensional space: complete positivity}

Complete positivity says, given the evolution equation (\ref{X5}), that $\langle w|{\cal R}'|w\rangle\geq0$ for an arbitrary $N^{2}$ dimensional vector $|w\rangle$  and for any initial density matrix ${\cal R}$ in the enlarged hilbert space.  We wish to prove  that complete positivity implies the eigenvalues are non-negative.  What we shall do is judiciously choose a single vector $|w\rangle$  and four pure density matrices ${\cal R}$ so that the expressions 
\[
 \langle w|{\cal R}'|w\rangle=\sum_{\alpha=1}^{N^{2}}\lambda^{\alpha}\langle w|{\bf E}^{\alpha}{\cal R}{\bf E}^{\alpha\dagger}|w\rangle,
\]
\noindent  are $\sim \lambda^{\beta}$, with a positive constant of proportionality. Therefore,  for complete positivity to hold, $ \lambda^{\beta}$ must be non-zero.  Here are choices that will do the job. 

    We shall choose the maximally entangled vector
\begin{equation}
|w\rangle\equiv\sum_{r=1}^{4}|\phi_{r}\rangle |\chi_{r}\rangle. \nonumber
\end{equation}
\noindent ($\langle w|w\rangle=4$, but it need not be normalized to 1). We construct the state vectors 
\[
|\psi\rangle^{\beta}\equiv\sum_{i,j=1}^{4}E_{ij}^{\dagger\beta}|\phi_{i}\rangle |\chi_{j}\rangle, 
\]
\noindent and  use them to make  four pure density matrices $|\psi\rangle^{\beta}\thinspace^{\beta}\negthinspace\langle\psi|$. (Note that $Tr|\psi\rangle^{\beta}\thinspace^{\beta}\negthinspace\langle\psi|=1$ because of the orthogonality relation Eq.(\ref{X4a})).  Then, for one $\beta$,   
\[
{\cal R}=\sum_{i,j,i',j'=1}^{4}E_{ij}^{\dagger\beta}E_{j'i'}^{\beta}|\phi_{i}\rangle |\chi_{j}\rangle\langle\phi_{i'}|\langle \chi_{j'}|. 
\]
\noindent  Putting this into Eq.(\ref{X5}), the complete positivity condition is 
\begin{eqnarray}
&&0\leq\langle w|{\cal R}'|w\rangle= \sum_{\alpha, r,r',i,j,i',j'=1}^{4}\lambda^{\alpha}E_{ij}^{\dagger\beta}E_{j'i'}^{\beta}\nonumber\\
&&\langle\phi_{r}|\langle \chi_{r}|{\bf E}^{\alpha}|\phi_{i}\rangle |\chi_{j}\rangle\langle\phi_{i'}|\langle \chi_{j'}| {\bf E}^{\dagger\alpha}
|\phi_{r'}\rangle |\chi_{r'}\rangle\nonumber\\
&&=\sum_{\alpha,i,j,i',j'=1}^{4}\lambda^{\alpha}E_{ij}^{\dagger\beta}E_{j'i'}^{\beta}E_{ji}^{\alpha}E_{'ij'}^{\dagger\alpha}\nonumber\\
&&=\sum_{\alpha=1}^{4}\lambda^{\alpha}Tr({\bf E}^{\dagger\beta}{\bf E}^{\alpha})Tr({\bf E}^{\beta}{\bf E}^{\alpha\dagger})\nonumber\\
&&=\sum_{\alpha=1}^{4}\lambda^{\alpha}
(\delta^{\alpha\beta})^{2}=\lambda^{\beta}\nonumber
\end{eqnarray}
\noindent (using the orthogonality relation (\ref{X4a})).

Thus, complete positivity implies $\lambda^{\beta}\geq 0$. 

\subsubsection{N-dimensional space: complete positivity}

We follow the same procedure in the N-dimensional case.  However, to be a bit more general, we shall  use an arbitrary  vector $|w\rangle$, and an 
arbitrary pure density matrix ${\cal R}$:
 \begin{subequations}
\begin{eqnarray}
|w\rangle&\equiv&\sum_{m,n=1}^{N}D_{mn}|\phi_{m}\rangle|\chi_{n}\rangle\label{X8a}\\
{\cal R}&\equiv&\sum_{k,l,k',l'=1}^{N}C_{kl}C_{k'l'}^{*}|\phi_{k}\rangle|\chi_{l}\rangle\langle\phi_{k'}|\langle\chi_{l'}|\label{X8b} 
\end{eqnarray}
\end{subequations}
\noindent where  $ C_{kl}$, $D_{mn}$ are yet to be specified complex constants.  The unit trace of ${\bf\rho}$ in Eq.(\ref{X8b}) requires  $Tr{\bf C^{\dagger}}{\bf C}=1$.  
Then, the complete positivity condition is
\begin{eqnarray}\label{X9}
0&\leq&\langle w|{\cal  R}'|w\rangle=\sum_{\alpha=1}^{N^{2}}\lambda^{\alpha}\langle w|{\bf E}^{\alpha}{\cal R}{\bf E}^{\alpha\dagger}|w\rangle\nonumber\\
&=&\sum_{\alpha=1}^{N^{2}}\lambda^{\alpha}\sum_{1}^{N}D_{m'n'}^{*}D_{mn}C_{kl}C_{k'l'}^{*}E_{m'k}^{\alpha}E_{m k'}^{\alpha*}\delta_{n'l}\delta_{l'n}\nonumber\\
&=&\sum_{\alpha=1}^{N^{2}}\lambda^{\alpha}Tr[{\bf C}{\bf D}^{\dagger} {\bf E}^{\alpha}]Tr[{\bf E}^{\alpha\dagger}{\bf D}{\bf C}^{\dagger}].
\end{eqnarray}
\noindent Now, choose ${\bf D}{\bf C}^{\dagger}={\bf E}^{\beta}$, for any particular $\beta$.  This choice can be made in many ways.  Two are ${\bf C}^{\dagger}={\bf E}^{\beta}$, ${\bf D}={\bf 1}$ (the choice made in the two-dimensional example just discussed) or 
$D={\bf E}^{\beta}$,  ${\bf C}=N^{-1/2}{\bf 1}$ (note, both choices respect $Tr{\bf C^{\dagger}}{\bf C}=1$). 
With this choice in Eq.(\ref{X9}), and with use of the orthonormality conditions Eq.(\ref{X4a}), we obtain as the consequence of complete positivity:
\begin{equation}\label{X10}
0\leq\sum_{\alpha=1}^{N^{2}}\lambda^{\alpha}(\delta^{\alpha\beta})^{2}=  \lambda^{\beta} \hbox{ for } 1\leq\beta\leq N^{2}. 
\end{equation}

 \section{Kraus Representation}
 
 We have now applied all the constraints needed to obtain a valid density matrix ${\bf\rho}'$ at a later time $t'$ from an earlier density matrix ${\bf\rho}$ at time $t$.  
This relation is  Eq.(\ref{X5}), supplemented by the orthonormality conditions (\ref{X4a}), the trace constraint (\ref{X6}) and 
the condition of non-negative eigenvalues (\ref{X10}).

 It is customary to define ${\bf M}^{\alpha}\equiv\sqrt{\lambda^{\alpha}}{\bf E}^{\alpha}$, so that Eqs.(\ref{X5}, \ref{X6}) can be written in terms of ${\bf M}^{\alpha}$ alone:
  \begin{subequations}
 \begin{eqnarray}
  &&{\bf \rho}'=\sum_{\alpha=1}^{N^{2}}{\bf M}^{\alpha}{\bf\rho} {\bf M}^{\alpha\dagger},\label{X11a}\\
&& \sum_{\alpha=1}^{N^{2}}{\bf M}^{\alpha\dagger}{\bf M}^{\alpha}={\bf 1}.\label{X11b}
\end{eqnarray}
 \end{subequations}
\noindent (However, the orthonormality conditions, written in terms of ${\bf M^{\alpha}}$,  now depend upon 
$\lambda^{\alpha}$).   Eq.(\ref{X11a}) is called the Kraus representation and $\{{\bf M^{\alpha}} \}$ are called Kraus operators\cite{Kraus}.  

We have proved the necessity of the Kraus representation, but it is  also sufficient.  That is,  for \textit{any}  $\{{\bf M^{\alpha}} \}$  satisfying 
Eqs.(\ref{X11a},\ref{X11b}), even for  more than $N^{2}$ operators, 
also with no orthonormality conditions imposed,  all the constraints on ${\bf\rho}'$ are satisfied.   It is easy to see  that hermiticity, trace 1 and positivity are satisfied.  Complete positivity requires a bit more work, and that is given in Appendix {\ref B}.  

This general statement of the Kraus representation might seem to imply a larger class than we have derived as necessary, but that is not so. Since the Kraus representation is hermitian, trace 1 and completely positive, it may  be written in the form Eq.(\ref{X5}), as we have shown. 

\section{Lindblad Equation}

	Now that we have satisfied all the constraints on the density matrix ${\bf\rho}'\equiv{\bf\rho}(t')$, we can  let $t'=t+dt$, and obtain the differential equation satisfied by ${\bf\rho}(t)$.	
For the rest of this paper we shall only treat the $N$-dimensional case since the argument is precisely identical for the two-dimensional case, except that $N=2$.  
\subsection{Eigenvectors and eigenvalues when $t'=t$}

First, lets see what we can say about the eigenvectors and eigenvalues when $t'=t$.   Then, Eq.(\ref{X5}) says
\begin{eqnarray}\label{15}
\rho_{ij}&=&\sum_{\alpha=1}^{N^{2}}\lambda^{\alpha}\sum_{r,s=1}^{N}E_{ir}^{\alpha}E_{sj}^{\alpha\dagger}\rho_{rs}\quad\hbox{or}\nonumber\\
0&=&\sum_{r,s=1}^{N}\Bigg[\sum_{\alpha=1}^{N^{2}}\lambda^{\alpha}E_{ir}^{\alpha}E_{sj}^{\alpha\dagger}-\delta_{ri}\delta_{js}\Bigg]\rho_{rs}.
\end{eqnarray}
\noindent As we have done before, successive replacement of ${\bf\rho}$ by the  $N^{2}$ members of the density matrix basis results in
\begin{equation}\label{16}
\delta_{ri}\delta_{js}=\sum_{\alpha=1}^{N^{2}}\lambda^{\alpha}E_{ir}^{\alpha}E_{sj}^{\alpha\dagger}.
\end{equation}
\noindent Multiply Eq.(\ref{16}) by $E_{js}^{\beta}$ and sum over $j,s$.  Use of the orthonormality relation (\ref{X4a}) gives 
\begin{equation}\label{17}
\delta_{ri}Tr{\bf E}^{\beta}=\lambda^{\beta}E_{ir}^{\beta}. 
\end{equation}

If $Tr{\bf E}^{\beta}\neq 0$ and $\lambda^{\beta}\neq0$,   Eq.(\ref{17}) says that all the eigenvectors are  $\sim{\bf 1}$.   But only one of a set of orthogonal eigenvectors can be proportional to the identity.  Therefore, for the rest of the eigenvectors,  $\lambda^{\beta}=0$ and $Tr{\bf E}^{\beta}=0$. 

Call one eigenvector ${\bf E}^{N^{2}}\equiv N^{-1/2}{\bf 1}$. From Eq.(\ref{17}), we find the associated eigenvalue $\lambda^{N^{2}}=N$. 

For $\beta\neq N^{2}$, the eigenvalues vanish.  Note that the condition $Tr{\bf E}^{\beta}{\bf 1}=0$ says that these eigenvectors are 
 orthogonal to ${\bf E}^{N^{2}}\sim {\bf 1}$. 
 
 And, indeed, in this case, Eq.(\ref{X5}) becomes the identity
\begin{equation}\label{18}
{\bf\rho}(t)={\bf 1}{\bf\rho}(t){\bf 1}=N\frac{1}{\sqrt{N}}{\bf 1}{\bf\rho}(t)\frac{1}{\sqrt{N}}{\bf 1}.
\end{equation} 

\subsection{Eigenvectors and eigenvalues when $t'=t+dt$}
	
	  When $t'=t+dt$,  the eigenvalues and eigenvectors change infinitesimally.  Accordingly we write
\begin{eqnarray}\label{19}
 &&\lambda^{N^{2}}(dt)=N[1-c^{N^{2}}dt], \medspace \lambda ^{\alpha}(dt)=c^{\alpha}dt \medspace(\alpha\neq N^{2}),\nonumber\\
 &&{\bf E}^{N^{2}}(dt)=\frac{1}{\sqrt{N}}[{\bf1}+{\bf B}dt], \medspace {\bf E}^{\alpha}(dt)= {\bf K}^{\alpha} \medspace(\alpha\neq N^{2}),\nonumber\\
\end{eqnarray}
 \noindent where the $c^{\alpha}$ are constants. We do not  include a term $\sim dt$ in the expression for ${\bf E}^{\alpha}(dt)$ since, because $\lambda ^{\alpha}(dt)\sim dt$, it would 
 contribute a negligible term $\sim (dt)^{2}$ to  Eqs.(\ref{X5},\ref{X6}).
 
 Because the eigenvalues must be positive, and because the eigenvalues sum to $N$ (equation following Eq.(\ref{X6})), we see that $c^{\alpha}\geq0$ (all $\alpha$).  
  ${\bf B}$ and ${\bf K}^{\alpha}$ are restricted by the orthonormality conditions, which we shall look at later.
  
  \subsection{The evolution equation}
  
  Putting Eqs.(\ref{19}) into the evolution equation (\ref{X5}) gives
 \begin{eqnarray}\label{20}
 &&{\bf \rho}(t+dt)=[1-c^{N^{2}}dt][{\bf 1}+{\bf B}dt]{\bf\rho}(t)[{\bf 1}+{\bf B^{\dagger}dt}]\nonumber\\
  &&+ dt\sum_{\alpha=1}^{N^{2}-1}c^{\alpha}{\bf K}^{\alpha}{\bf\rho}(t){\bf K}^{\alpha\dagger}, \medspace\hbox {or in the limit}\medspace dt\rightarrow 0,\nonumber \\
  &&\frac{d}{dt}{\bf\rho}(t)=-c^{N^{2}}{\bf\rho}(t)+{\bf B}{\bf\rho}(t)+{\bf\rho}(t){\bf B^{\dagger}}+\sum_{\alpha=1}^{N^{2}-1}c^{\alpha}{\bf K}^{\alpha}{\bf\rho}(t){\bf K}^{\alpha\dagger}.\nonumber \\ 
 \end{eqnarray}
   Putting  Eqs.(\ref{19}) into the trace constraint (\ref{X6}) gives
   \begin{eqnarray}\label{21}
 &&N[1-c^{N^{2}}dt][{\bf 1}+{\bf B^{\dagger}dt][{\bf 1}+{\bf B}}dt]\nonumber\\
&&\qquad\qquad\qquad+dt\sum_{\alpha=1}^{N^{2}-1}c^{\alpha}{\bf K}^{\alpha\dagger}{\bf K}^{\alpha}={\bf 1},\medspace\hbox{or,}\nonumber\\
&&c^{N^{2}}{\bf 1}= {\bf B}+{\bf B}^{\dagger}+\sum_{\alpha=1}^{N^{2}-1}c^{\alpha}{\bf K}^{\alpha\dagger}{\bf K}^{\alpha}.
\end{eqnarray}
\noindent Using (\ref{21}) to replace $c^{N^{2}}$ in (\ref{20}) (specifically, $c^{N^{2}}\rho=(1/2)[c^{N^{2}}{\bf 1}{\bf\rho}+{\bf\rho}c^{N^{2}}{\bf 1}]$) results in 
 \begin{eqnarray}\label{22}
&&\frac{d}{dt}{\bf\rho}(t)=\Big[\frac{1}{2}({\bf B}-{\bf B}^{\dagger}), {\bf\rho}(t)\Big]\nonumber\\
&&-\frac{1}{2}\sum_{\alpha=1}^{3}c^{\alpha}[{\bf K}^{\alpha}{\bf K}^{\alpha\dagger} {\bf\rho}(t)+ {\bf\rho}(t){\bf K}^{\alpha}{\bf K}^{\alpha\dagger}-{\bf K}^{\alpha}{\bf\rho}(t){\bf K}^{\alpha\dagger}].\nonumber\\
 \end{eqnarray}
 
 \subsection{The Lindblad equation}
 
If we define $-i{\bf H}\equiv (1/2)({\bf B^{\dagger}}-{\bf B})$ and 
${\bf L}^{\alpha}=\sqrt{c^{\alpha}}\bf K^{\alpha}$,  the evolution equation (\ref{22}) becomes  the Lindblad equation (\ref{Z3}):
\begin{eqnarray}\label{23}
&&\frac{d}{dt}{\bf\rho}(t)=-i\Big[{\bf H}, {\bf\rho}(t)\Big]\nonumber\\
&&-\frac{1}{2}\sum_{\alpha=1}^{N^{2}-1}[{\bf L}^{\alpha}{\bf L}^{\alpha\dagger} {\bf\rho}(t)+ {\bf\rho}(t){\bf L}^{\alpha}{\bf L}^{\alpha\dagger}-2{\bf L}^{\alpha}{\bf\rho}(t){\bf L}^{\alpha\dagger}].\nonumber\\
 \end{eqnarray}

 \subsection{Orthonormality conditions}\label{VIE}

It is a consequence of this derivation that the Lindblad operators ${\bf L}^{\alpha}$ in Eq.(\ref{23}) are not arbitrary operators, because they are restricted by the orthonormality conditions (\ref{X4a}). 
Putting Eqs.(\ref{19}) into Eq.(\ref{X4a}) constrains  ${\bf B}$,  ${\bf K}^{\alpha}$. 

 For what follows, we recall from the discussion in section II that  $\{\bf K^{\alpha}\}$ can be regarded in two ways.  In one way, they are regarded as $N^{2}-1$ operators acting on vectors in an $N$ dimensional space, with matrix elements $ K_{ij}^{\alpha}$ ($i,j=1 ... N$). In the other way, they are regarded as $N^{2}-1$ vectors in an $N^{2}$ dimensional space, each with components ($K_{11}^{\alpha}, K_{12}^{\alpha},  ...  K_{NN}^{\alpha}$). In particular, the trace of two operators is the same as the scalar product of two vectors, as in Eq.(\ref{X4a}) 
 
 The orthonormality relation (\ref{X4a}), applied successively to $(\alpha=\beta=N^{2})$,  $(\alpha\neq N^{2},  \beta=N^{2})$, $(\alpha\neq N^{2},  \beta\neq N^{2})$, with 
 use of Eqs.(\ref{19}), are
 
 \begin{subequations}\label{24}
\begin{eqnarray}
&&Tr[{\bf B}+{\bf B}^{\dagger}]=0,\label{24a}\\
&&Tr{\bf K}^{\alpha}=0,\label{24b}, \medspace(\alpha=1, ... N^{2}-1)\\
&&Tr {\bf K}^{\alpha}{\bf K}^{\beta\dagger}=\delta^{\alpha\beta} \medspace(\alpha, \beta=1, ... N^{2}-1).\label{24c}
\end{eqnarray}
\end{subequations}

Eq.(\ref{24a}) says that the hermitian part of  ${\bf B}$ vanishes.  This provides no restriction at all on ${\bf H}$, which is the anti-hermitian part of ${\bf B}$.

Eq.(\ref{24c}) says that the vectors ${\bf K}^{\alpha}$ are orthonormal.

Eq.(\ref{24b}) says that  $Tr{\bf K}^{\alpha}{\bf 1}=0$, which implies that $N^{-1/2}{\bf 1}$ completes the orthonormal set.

 \subsection{The general Lindblad form}\label{VIF}

 We shall now show that the Lindblad equation (\ref{23}) with \textit{arbitrary} Lindblad operators (no constraints whatsoever)  can be transformed to new, \textit{constrained},  Lindblad operators of Eq.(\ref{22}) by adding a constant (achieving the vanishing trace constraint  (\ref{24b})) followed by a unitary transformation(achieving the orthogonality constraint  (\ref{24c})).

First, we see that we can transform the arbitrary Lindblad operators ${\bf L}^{\alpha}$ to Lindblad operators ${\bf L}'^{\alpha}$ which are traceless.   Define ${\bf L}^{\alpha}\equiv{\bf L}^{'\alpha}+k^{\alpha}{\bf 1}$, where the $k^{\alpha}$ are $N^{2}-1$ constants,  and substitute that into the Lindblad equation, obtaining :
 \begin{eqnarray}
&&\frac{d}{dt}{\bf\rho}(t)=-i\Big[{\bf H}+i(k^{\alpha}{\bf L}'^{\alpha\dagger}-k^{\alpha*}{\bf L}'^{\alpha}), {\bf\rho}(t)\Big]\nonumber\\
&&-\frac{1}{2}\sum_{\alpha=1}^{3}[{\bf L}'^{\alpha}{\bf L}'^{\alpha\dagger} {\bf\rho}(t)+ {\bf\rho}(t){\bf L}'^{\alpha}{\bf L}'^{\alpha\dagger}-2{\bf L}'^{\alpha}{\bf\rho}(t){\bf L}'^{\alpha\dagger}].\nonumber
 \end{eqnarray}
\noindent With a redefinition of ${\bf H}$, this is again the Lindblad equation, expressed in terms of ${\bf L}'^{\alpha}$. By choosing $k^{\alpha}=N^{-1}Tr{\bf L}^{\alpha}$, the new Lindblad operators satisfy $Tr{\bf L}'^{\alpha}=0$.  

Now write the $N^{2}-1$ Lindblad operators ${\bf L}^{\alpha}$  (hereafter assumed traceless) in terms of $N^{2}-1$ new operators $\tilde{\bf L}^{\beta}$ ($\alpha$, $\beta =1,... N^{2}-1$) using the linear transformation
\begin{equation}\label{25}
{\bf L}^{\alpha}=\sum_{\beta=1}^{N^{2}-1}U^{\alpha, \beta}\tilde{\bf L}^{\beta}.
\end{equation}
\noindent We ask what the matrix $U^{\alpha, \beta}$ must be in order that the Lindblad form be unchanged ($\tilde{\bf L}^{\alpha}$ replacing ${\bf L}^{\alpha}$  
in Eq.(\ref{23})).  With arbitrary  operators ${\bf A}_{1}$, ${\bf A}_{2}$, ${\bf A}_{3}$,  
\[
\sum_{\alpha=1}^{N^{2}-1}{\bf A}_{1}L^{\alpha}{\bf A}_{2}L^{\alpha\dagger}{\bf A}_{3}=\sum_{\alpha,\beta, \beta'=1}^{N^{2}-1}U^{\alpha, \beta}U^{*\alpha, \beta'}{\bf A}_{1}\tilde{\bf L}^{\beta}{\bf A}_{2}
\tilde{\bf  L}^{\beta'\dagger}{\bf A}_{3}.
\]
\noindent (The three terms in the Lindblad equation have two of the ${\bf A}={\bf 1}$ while the third  ${\bf A}={\bf \rho}$).  This  equals $\sum_{\beta=1}^{N^{2}-1}{\bf A}_{1}\tilde{\bf L}^{\beta}{\bf A}_{2}\tilde{\bf L}^{\beta\dagger}{\bf A}_{3}$, leaving  
the Lindblad equation unchanged in form, if and only if the  matrix $U^{\alpha, \beta}$ is unitary, $\sum_{\alpha=1}^{N^{2}-1}U^{\dagger\beta', \alpha}U^{\alpha, \beta}=\delta^{\beta,\beta'}$.

Inverting Eq.(\ref{25}), we see that $\tilde{\bf L}^{\beta}$ is traceless. 

Now, consider the matrix $Tr{\bf L}^{\alpha}{\bf L}^{\alpha'\dagger}$.  It is hermitian, so its eigenvalues are real and its eigenvectors are orthogonal.  It can be brought to diagonal form by properly choosing our unitary transformation, so we obtain
\begin{equation}\label{26}
Tr \tilde{\bf L}^{\beta}\tilde{\bf L}^{\beta'\dagger}=\sum_{\alpha=1}^{N^{2}-1}U^{\dagger\beta, \alpha} U^{\alpha', \beta}Tr{\bf L}^{\alpha}{\bf L}^{\alpha'\dagger}=\tilde c^{\beta}\delta^{\beta\beta'}.
\end{equation}
This is almost the orthogonality constraint  (\ref{24c}).  The eigenvalues $\tilde c^{\beta}$ are non-negative, since it follows from Eq.(\ref{26}) that 
\[
\tilde c^{\beta}=Tr \tilde{\bf L}^{\beta}\tilde{\bf L}^{\beta\dagger}= \sum_{i,j=1}^{N}L_{ij}^{\beta}L_{ij}^{*\beta}\geq 0.
\]
Thus, according to Eq.(\ref{26}), the $\tilde{\bf L}^{\beta}$  are orthogonal vectors, 
with squared norm $\tilde c^{\beta}$.  We can add one more vector $\sim {\bf 1}$ to complete the set, orthogonal to the rest since $Tr \tilde{\bf L}^{\beta}{\bf 1}=Tr \tilde{\bf L}^{\beta}=0.$

We can define  new operators $\tilde{\bf K}^{\beta}$ which are orthonormal and traceless, by $\tilde{\bf L}^{\beta}\equiv\sqrt{\tilde c^{\beta}}\tilde{\bf K}^{\beta}$.  In terms of these operators, the Lindblad equation (\ref{23}) written in terms of $\tilde{\bf L}^{\beta}$ becomes
\begin{eqnarray}\label{27}
&&\frac{d}{dt}{\bf\rho}(t)=-i\Big[{\bf H}, {\bf\rho}(t)\Big]\nonumber\\
&&-\frac{1}{2}\sum_{\alpha=1}^{N^{2}-1}\tilde c^{\alpha}[\tilde{\bf K}^{\alpha}\tilde{\bf K}^{\alpha\dagger} {\bf\rho}(t)+ {\bf\rho}(t)\tilde{\bf K}^{\alpha}\tilde{\bf K}^{\alpha\dagger}-2\tilde{\bf K}^{\alpha}{\bf\rho}(t)\tilde{\bf K}^{\alpha\dagger}].\nonumber\\
 \end{eqnarray}
\noindent This is precisely Eq.(\ref{22}) with $\tilde{\bf K}^{\alpha}$, $\tilde c^{\alpha}$, replacing ${\bf K}^{\alpha}$, $\ c^{\alpha}$.  Moreover, the orthonormality constraints Eqs.(\ref{24b},\ref{24c}) on ${\bf K}^{\alpha}$ are satisfied by $\tilde{\bf K}^{\alpha}$.  

\subsection{Concluding remarks}
 
 We have shown that the constraints on the density matrix ${\bf \rho}'$ mandate its evolution equation (\ref{22}), where $c^{\alpha}\geq 0$ and ${\bf B},{\bf K}^{\alpha}$ satisfy the 
 constraints (\ref{24}).  We then showed that this is completely equivalent to the Lindblad Eq.(\ref{23}), with no constraints at all on the $N^{2}-1$ Lindblad 
 operators ${\bf L}^{\alpha}$.  
 
 However, there is no need to restrict the Lindblad equation to no more than $N^{2}-1$ operators.  We conclude our presentation by showing  that the Lindblad equation with any number of operators has all the required properties. (Of course, from what we have shown,  such an equation may be reduced to one with no more than $N^{2}-1$ operators.)
 
 Looking at the Lindblad equation in that case,
 \begin{eqnarray}\label{28}
&&\frac{d}{dt}{\bf\rho}(t)=-i\Big[{\bf H}, {\bf\rho}(t)\Big]\nonumber\\
&&-\frac{1}{2}\sum_{\alpha}[{\bf L}^{\alpha}{\bf L}^{\alpha\dagger} {\bf\rho}(t)+ {\bf\rho}(t){\bf L}^{\alpha}{\bf L}^{\alpha\dagger}-2{\bf L}^{\alpha}{\bf\rho}(t){\bf L}^{\alpha\dagger}], 
 \end{eqnarray}
 \noindent   it is easy to see  that hermiticity, and trace 1 (in the form $dTr{\bf\rho}(t)/dt=0$, with $Tr{\bf\rho}(t_{0})={\bf 1}$) are satisfied.  Complete positivity requires a bit more work. 

We have shown in Appendix {\ref B} that complete positivity holds for the  Kraus form (\ref{X11a}) subject to the trace constraint (\ref{X11b}), with an arbitrary number of operators.  So, if Eq.(\ref{28}) can be written in the Kraus form with the trace constraint, we have shown it is completely positive. 

Accordingly, we write  Eq.(\ref{28}) as 
\begin{eqnarray}\label{29}
&&{\bf\rho}(t+dt)=\Big[{\bf 1}-dt(i{\bf H}+\frac{1}{2}\sum_{\alpha}{\bf L}^{\alpha\dagger}{\bf L}^{\alpha})\Big]{\bf\rho}(t)\nonumber\\
&&\cdot\Big[{\bf 1}-dt(-i{\bf H}+\frac{1}{2}\sum_{\alpha}{\bf L}^{\alpha\dagger}{\bf L}^{\alpha})\Big]+dt\sum_{\alpha}{\bf L}^{\alpha}{\bf\rho}(t){\bf L}^{\alpha\dagger}.\nonumber\\
\end{eqnarray}
 \noindent Identifying the Kraus operators as
\begin{equation}\label{30}
{\bf M}^{0}=[{\bf 1}-dt(i{\bf H}+\frac{1}{2}\sum_{\alpha=1}^{N^{2}-1}{\bf L}^{\alpha\dagger}{\bf L}^{\alpha})\Big], {\bf M}^{\alpha\neq0}= \sqrt{dt}{\bf L}^{\alpha},
\end{equation}
 \noindent gives the Kraus form Eq.(\ref{X11a}), 
 \begin{equation}\label{31}
{\bf\rho}(t+dt)=\sum_{\alpha}{\bf M}^{\alpha}{\bf\rho}(t){\bf M}^{\alpha\dagger}.
\end{equation} 

Now, take the trace of Eq.(\ref{31}). Since Eq.(\ref{28}) implies $Tr{\bf\rho}(t+dt)=1=Tr{\bf 1}{\bf\rho}(t)$, the result is
\begin{equation}\label{32}
1=Tr\sum_{\alpha}{\bf M}^{\alpha\dagger}{\bf M}^{\alpha}{\bf\rho}(t) \hbox{ or }Tr\Big[\sum_{\alpha}{\bf M}^{\alpha\dagger}{\bf M}^{\alpha}-{\bf 1}\Big]{\bf\rho}(t).
\end{equation}
\noindent As we have done before,  by successively replacing ${\bf\rho}(t)$ by the members of the density matrix basis, we obtain the Kraus trace constraint (\ref{X11b}):
\begin{equation}\label{32}
Tr\sum_{\alpha}{\bf M}^{\alpha\dagger}{\bf M}^{\alpha}={\bf 1}.
\end{equation}
 
 Therefore, the Lindblad form with an arbitrary number of Lindblad operators, is completely positive. 
 
 \begin{appendix}
 \section{A Class of Positive Density Matrices With Negative Superoperator Eigenvalues}\label{A}
 Consider  the Pauli+1 matrices (multiplied by $1/\sqrt{2}$) as the eigenvectors ${\bf E}^{\beta}$. As noted in Section IV, they satisfy the trace constraint (\ref{X6}) if $\sum_{\alpha=1}^{4}\lambda^{\alpha}=2$. 
  Eq.(\ref{X5}) becomes 
\begin{eqnarray}\label{A1}
&&{\bf \rho}'
=\sum_{\alpha=1}^{4}\lambda^{\alpha}{\bf E}^{\alpha}{\bf\rho} {\bf E}^{\alpha\dagger}\nonumber\\
&&=\frac{1}{2}[\lambda^{1}{\bf \sigma}^{1}{\bf\rho} {\bf \sigma}^{1}+\lambda^{2}{\bf \sigma}^{2}{\bf\rho} {\bf \sigma}^{2}+\lambda^{3}{\bf \sigma}^{3}{\bf\rho} {\bf \sigma}^{3}+\lambda^{4}{\bf1}{\bf\rho} {\bf 1}]\nonumber\\
&&\negmedspace\negmedspace\negmedspace\negmedspace=\frac{1}{2}\begin{bmatrix}(\lambda^{1}+\lambda^{2})\rho_{22}+(\lambda^{3}+\lambda^{4})\rho_{11},&(\lambda^{1}-\lambda^{2})\rho_{21}-(\lambda^{3}-\lambda^{4})\rho_{12}
\\(\lambda^{1}-\lambda^{2})\rho_{12}-(\lambda^{3}-\lambda^{4})\rho_{21},&(\lambda^{1}+\lambda^{2})\rho_{11}+(\lambda^{3}+\lambda^{4})\rho_{22}
\end{bmatrix}.\nonumber\\
\end{eqnarray}
\noindent 
 \noindent Now, we can successively replace ${\bf \rho}$  by the four density basis matrices, and demand that the $\lambda^{\alpha}$ be chosen so ${\bf \rho}'$ is positive for all. Since the sum of these four ${\bf \rho}$'s with positive coefficients (adding up to 1) is the  most general two-dimensional density matrix, then the most general ${\bf \rho}'$ will be positive. 
 
 The first density basis matrix has $\rho_{11}=1$ and the rest of the matrix elements vanishing. Then, 
 \begin{equation}\label{A2}
{\bf \rho}'
=\frac{1}{2}\begin{bmatrix}(\lambda^{3}+\lambda^{4})&0
\\0&(\lambda^{1}+\lambda^{2})
\end{bmatrix}.
\end{equation}
Since the eigenvalues of the density matrix must lie between 1 and 0,  we obtain the two conditions:
 \begin{subequations}\label{A3}
 \begin{eqnarray}
&&2\geq\lambda^{1}+\lambda^{2}\geq 0\label{A3a}\\
&&2\geq\lambda^{3}+\lambda^{4}\geq 0.\label{A3b}
\end{eqnarray}
\end{subequations}
\noindent The second density basis matrix,  with $\rho_{22}=1$ and the rest of the matrix elements vanishing, gives the same results.

The third density basis matrix is ${\bf\rho}=(1/2)[{\bf 1}+{\bf \sigma}_{1}]$.  Using $\sum_{\alpha=1}^{4}\lambda^{\alpha}=2$ to simplify the result, we obtain:
\begin{equation}\label{A4}
{\bf \rho}'
=\frac{1}{2}\begin{bmatrix}1&(\lambda^{1}+\lambda^{4}-1)\\
(\lambda^{1}+\lambda^{4}-1)&1
\end{bmatrix}.
\end{equation}
The eigenvalues of ${\bf \rho}'$  here are $(\lambda^{1}+\lambda^{4})/2$, $1-(\lambda^{1}+\lambda^{4})/2$, so the condition that they lie between 0 and 1 is
 \begin{equation}\label{A5}
2\geq\lambda^{1}+\lambda^{4}\geq 0\\.
\end{equation}

The fourth density basis matrix is ${\bf\rho}=(1/2)[{\bf 1}+{\bf \sigma}_{2}]$.  Using $\sum_{\alpha=1}^{4}\lambda^{\alpha}=2$ to simplify the result, we obtain:
\begin{equation}\label{A6}
{\bf \rho}'
=\frac{1}{2}\begin{bmatrix}1&i(\lambda^{2}+\lambda^{4}-1)\\
-i(\lambda^{2}+\lambda^{4}-1)&1
\end{bmatrix}.
\end{equation}
The eigenvalues of ${\bf \rho}'$  here are $(\lambda^{2}+\lambda^{4})/2$, $1-(\lambda^{2}+\lambda^{4})/2$, so the condition that they lie between 0 and 1 is
 \begin{equation}\label{A7}
2\geq\lambda^{2}+\lambda^{4}\geq 0\\.
\end{equation}

So, we have obtained the result that ${\bf \rho}'$ will  be positive if Eqs.(\ref{A3a}, \ref{A3b},\ref{A5},\ref{A7}) and 
\begin{equation}\label{A8}
\sum_{\alpha=1}^{4}\lambda^{\alpha}=2
\end{equation}
\noindent  are satisfied.  The sum of  Eqs.(A3b, \ref{A5}, \ref{A7}))minus (\ref{A8}) tells us that $2\geq \lambda^{4}\geq -1$. 

 Eq.(\ref{A8}) and the constraint boundaries  are  three-dimensional hyperplanes in the four dimensional $\lambda$-space.
  Their intersections delineate the allowed areas for the eigenvalues.  
 
 We shall be content here to set $\lambda^{2}=\lambda^{1}$ (in which case 
 Eqs.(A3a) simplifies to $1\geq \lambda^{1}\geq 0$). Then, 
 Eq.(\ref{A8}) describes a plane in $\{\lambda^{1},\lambda^{3}, \lambda^{4}\}$ space, and its intersection with the 
 constraint boundary planes can be drawn.  This is shown in Fig. 1. There are two regions where one of the eigenvalues is negative and the other two are positive:  the points in the heavily outlined upper left triangle have $\lambda^{3}\leq 0$, and 
 the points in the heavily outlined lower right triangle have $\lambda^{4}\leq 0$.
 
 \begin{figure}[h]
\begin{center}
\includegraphics[width=0.5\textwidth]{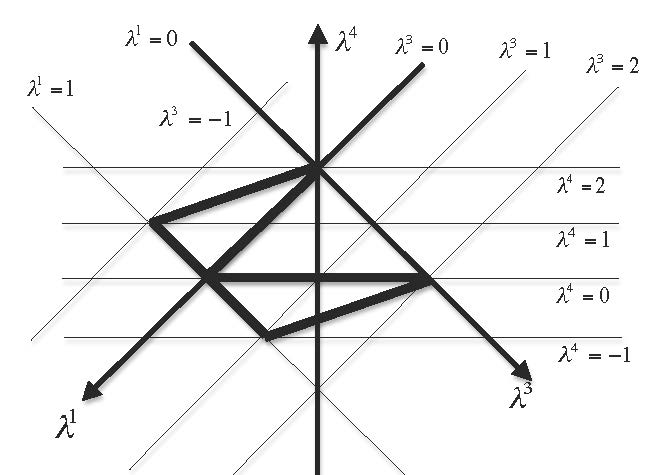}
\caption{\label{f1} Allowed regions of eigenvalues for a positive density matrix ${\bf \rho}'$. The two dark-outlined triangular regions are where an eigenvalue is negative.  They abut 
an isosceles triangle, the  restricted region of complete positivity, where all the eigenvalues are positive.}
\label{default}
\end{center}
\end{figure} 
 
 \section{Complete Positivity of the Kraus representation}\label{B}
The Kraus form Eq.(\ref{X11a}) and the Kraus constraint Eq.(\ref{X11b}), generalized to \textit{any} number of \textit{arbitrary} operators ${\bf M}^{\alpha}$, are respectively
 \begin{subequations}
 \begin{eqnarray}
  &&{\bf \rho}'=\sum_{\alpha}{\bf M}^{\alpha}{\bf\rho} {\bf M}^{\alpha\dagger},\label{B1a}\\
&& \sum_{\alpha}{\bf M}^{\alpha\dagger}{\bf M}^{\alpha}={\bf 1}.\label{B1b}
\end{eqnarray}
 \end{subequations}
\noindent We want to show complete positivity.  Call any one of the ${\bf M}^{\alpha}\equiv {\bf M}$.  If we can show complete positivity for 
${\bf M}{\bf \rho}{\bf M}^{\dagger}$ for an arbitrary ${\bf M}$, then Eq.(\ref{B1a}), which involves a sum of such terms, will be completely positive. And, it is only necessary to prove complete positivity for ${\cal R}={\bf\rho}\times \tilde{\bf\rho}^\mu$, where $\tilde{\bf\rho}^\mu$ is any possible  basis density matrix (described in the paragraph following  Eq.(\ref{X3})) in the added hilbert space, since the most general density matrix in the direct product hilbert space is the linear sum of such terms. 

We now calculate
\begin{equation}\label{B2}
\langle v|{\cal R}'|v\rangle=\langle v|{\bf M}{\bf\rho}{\bf M}^{\dagger}\times\tilde{\bf\rho}^{\mu}|v\rangle.
\end{equation}
\noindent for arbitrary $|v\rangle=\sum_{m,n=1}^{N}D_{mn}|\phi_{m}\rangle|\chi_{n}\rangle$.  There is no loss of generality if we pick the basis vectors  $|\chi_{n}\rangle$ in the added hilbert space any way we like.  We shall pick them  to be the eigenstates of  
$\tilde{\bf\rho}^{\mu}$. Now, we note that each $\tilde{\bf\rho}^{\mu}$ has one eigenvalue 1 and the remaining $N-1$ eigenvalues are 0.  
 Call the eigenvector  $|\chi_{1}\rangle$ which corresponds to the eigenvalue 1.  Then, 
\begin{eqnarray}\label{B3}
&&\langle v|{\bf M}{\bf \rho}{\bf M}^{\dagger}\times\tilde{\bf\rho}^{\mu}|v\rangle=\sum_{m,m' =1}^{N}
D_{m'1}^{*}D_{m1}\langle \phi_{m'}|{\bf M}{\bf \rho}{\bf M}^{\dagger}|\phi_{m}\rangle\nonumber\\
&&=\Big[\sum_{m' =1}^{N}D_{m'1}^{*}\langle \phi_{m'}|{\bf M}\Big]{\bf\rho}
\Big[\sum_{m' =1}^{N}D_{m1}{\bf M}^{\dagger}| \phi_{m}\rangle\Big]\geq 0,
\end{eqnarray}
\noindent Therefore,  the Kraus form with arbitrary ${\bf M}^{\alpha}$'s is completely positive.

\end{appendix}

 \end{document}